\documentstyle[twoside,fleqn,espcrc2,epsfig,amssymb]{article}

\newcommand{\half}{\mbox{\small $\frac{1}{2}$}}          
\newcommand{\qchpt}{$q\chi PT$}

\def\lsim{\mathrel{\rlap{\lower4pt\hbox{\hskip1pt$\sim$}}
    \raise1pt\hbox{$<$}}}                
\def\gsim{\mathrel{\rlap{\lower4pt\hbox{\hskip1pt$\sim$}}
    \raise1pt\hbox{$>$}}}                

\def\3{\ss}

\newcommand{\AmS}{{\protect\the\textfont2
  A\kern-.1667em\lower.5ex\hbox{M}\kern-.125emS}}

\title{
       \vspace{-4.00cm}                                     
       {\normalsize DESY 99--131}    \\[-0.2cm]             
       {\normalsize HLRZ 99--40}     \\[-0.2cm]             
       {\normalsize FUB-HEP/99--5}   \\[-0.2cm]             
       {\normalsize HUB--EP--99/53}  \\[-0.2cm]             
       {\normalsize TPR-99-17}       \\[-0.2cm]             
       {\normalsize September 1999}  \\                     
       \vspace{1.42cm}                                      
       Hadron masses and decay constants in quenched QCD%
            \thanks{Talk given by D.~Pleiter at Lattice '99,      
                    Pisa, Italy.}}                       

\author{M.~G\"ockeler%
           \address{Institut f\"ur Theoretische Physik, Universit\"at
                    Regensburg, D-93040 Regensburg, Germany},
        R.~Horsley%
           \address{Institut f\"ur Physik, Humboldt-Universit\"at zu Berlin,
                    D-10115 Berlin, Germany},
        D.~Petters%
	   \address{Institut f\"ur Theoretische Physik,
	            Freie Universit\"at Berlin, D-14195 Berlin, Germany}%
	   $^{\rm ,}$%
           \address{Deutsches Elektronen-Synchrotron DESY and NIC,
                    D-15735 Zeuthen, Germany},
        D.~Pleiter$^{\rm c,d}$,
        P.E.L.~Rakow$^{\rm a}$,
        G.~Schierholz$^{\rm d,}$%
           \address{Deutsches Elektronen-Synchrotron DESY,
                    D-22603 Hamburg, Germany}
	and
        P.~Stephenson%
           \address{Dipartimento di Fisica,
                    Universit\`a degli Studi di Pisa e INFN,
                    Sezione di Pisa, 56100 Pisa, Italy}
	(QCDSF Collaboration)
}
       
\begin{document}

\begin{abstract}
We present results for the mass spectrum and decay constants using
non-perturbatively O(a) improved Wilson fermions.
Three values of $\beta$ and 30 different quark masses are used
to obtain the chiral and continuum limits.
Special emphasis will be given to the question of
taking the chiral limit and
the existence of non-analytic behavior
predicted
by quenched chiral perturbation theory ({\qchpt}).
\vspace*{-0.3cm}
\end{abstract}

\maketitle

\setcounter{footnote}{0}


\section{INTRODUCTION}

The chiral and continuum extrapolations of lattice results
in quenched QCD
have become an important issue since more accurate data for small
quark masses and results for different lattice spacings have become
available. Our simulations
have been done for three different values of
the gauge coupling, $\beta=6.0$, $6.2$ and $6.4$ ($a \approx
0.09-0.05$ fm), and on lattices of size $1.5-2.3$ fm. Our propagators
have been generated for 7-12 different values of the hopping parameter
$\kappa$ with
the ratio of pseudoscalar and vector meson mass
$m_{PS}/m_V$ in the range of $0.41-0.98$.

\section{CHIRAL EXTRAPOLATION}

Based on small quark mass expansions, chiral perturbation theory
can give information about how the extrapolation
to the chiral limit has to be done.
For quenched QCD,
{\qchpt} predicts non-analytic terms, e.g.~for the pseudoscalar mass
it gives to one-loop~\cite{Sharpe:1993gr}
\begin{equation}
m_{PS}^2 / \tilde{m}_q =
C_0 \tilde{m}_q^{-\frac{\delta}{1+\delta}} + C_1 \tilde{m}_q + O(\tilde{m}_q^2).
\label{eq:chiral_mpivsmq}
\end{equation}
For full QCD, the parameter $\delta$ should be zero.

In order to determine $\delta$ we fitted our data using the ansatz
\begin{eqnarray}
\lefteqn{a m_{PS}^2 / \tilde{m}_q^{R} = C_\beta\;
    \left[ (a \tilde{m}_q^{R})^{-\frac{\delta}{1+\delta}}
           + A_{1,\beta}\;a \tilde{m}_q^{R} \right.} \nonumber\\
&&  \left. +\; A_{2,\beta}\;(a \tilde{m}_q^{R})^2 \right]
    / \left[ 1 + B_{1,\beta}\;a \tilde{m}_q^{R} \right],
\label{eq:ansatz_mpivsmq}
\end{eqnarray}
where
$\tilde{m}_q^{R} = Z_M(g_0) \frac{1+b_A am_q}{1+b_P am_q}\,\tilde{m}_q$
is the renormalization group invariant, and
\begin{figure}[h]
\vspace*{-0.8cm}
\includegraphics[width=6.5cm]{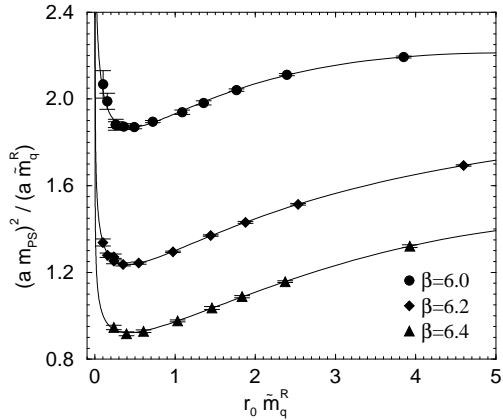}
\vspace*{-1.1cm}
\caption{The ratio $m_{PS}^2 / \tilde{m}_q^R$ as a function of $\tilde{m}_q^R$.
         The solid line is a fit using eq.~(\ref{eq:ansatz_mpivsmq}).}
\vspace*{-0.1cm}
\label{fig:chiral_mpivsmq}
\end{figure}
$\tilde{m}_q$ the bare Ward identity quark mass.
The non-perturbative renormalization constant and the perturbative
improvement coefficients $b_A$ and $b_P$ have been computed
in \cite{Sint:1997jx_Capitani:1998mq}.
According to this ansatz
$m_{PS}^2 / \tilde{m}_q^R$ should be linear up to heavier quark
masses. Our results are shown in Fig.~\ref{fig:chiral_mpivsmq}. From this
fit we get $\delta \approx 0.14(2)$.

\begin{figure}[t]
\vspace*{0.2cm}
\includegraphics[width=6.4cm]{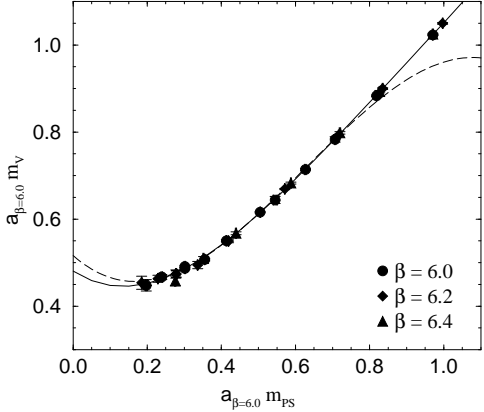}
\vspace*{-1.1cm}
\caption{Vector meson mass as a function of the pseudoscalar mass.
         The solid line is a fit using eq.~(\ref{eq:ansatz_pade})
         with $N=4$. The dashed line is a fit using a
         cubic polynomial with $a m_{PS}<0.6$.}
\vspace*{-0.6cm}
\label{fig:chiral_rho-lin}
\end{figure}

For the vector meson mass $m_V$ and the nucleon mass $m_N$,
{\qchpt} predicts~\cite{Booth:1997hk,Labrenz:1996jy}
\begin{eqnarray}
\label{eq:fit_mvn}
\lefteqn{a m_{V,N} = a M_{V,N} + C_{1/2}\, a m_{PS}}   \\
&& + C_1\, (a m_{PS})^2 + O\left((a m_{PS})^3\right) \nonumber
\end{eqnarray}
with $C_{1/2} \approx -1.5$ for $m_V$ and
$C_{1/2} = - (3\pi/2)\,(D-3F)^2\,\delta \approx -0.24$ for the
nucleon (D and F taken from \cite{Jaffe:1990jz}).
For full QCD, $C_{1/2}$ should be zero.

To reduce the number of free parameters in our fits, we make the assumption
that the coefficients $C_{1/2}$, $C_1$, ... do not show any scaling
violations, i.e.~do not depend on $\beta$.
To take the different scales into account we
replace in eq.~(\ref{eq:fit_mvn}) $a \rightarrow \tilde{a}_\beta =
a_{\beta=6.0} s_\beta$ and use the scale parameter
$s_\beta$ as an additional
fit variable.
We fitted our data using a Pad\'e-like ansatz,
\begin{equation}
\tilde{a}_\beta m_{V,N} =
  \frac{\tilde{a}_\beta M_{V,N}
        + \sum_{i=1}^N A_i \left(\tilde{a}_\beta m_{PS}\right)^i}
       {1 + B_{N-1} \left(\tilde{a}_\beta m_{PS}\right)^{N-1}},
\label{eq:ansatz_pade}
\end{equation}
which is again chosen to give the correct heavy quark limit,
and a polynomial ansatz, which corresponds to Eq.~\ref{eq:ansatz_pade}
with $B_{N-1}=0$. For the Pad\'e-like ansatz
the results for $M_{V,N}$ are close to constant for
$0.1<B_{N-1}<1.5$, we therefore use $B_{N-1} = 1$.
For $m_V$ we find $C_{1/2} = -0.6(1)$ and $-0.7(3)$ for
the Pad\'e-like and the polynomial ansatz, respectively. The results are
plotted in Fig.~\ref{fig:chiral_rho-lin}. For $m_N$ we find
$C_{1/2} = 0.8(8)$ and $0.4(5)$.
Our results for $s_\beta$ deviate at most 2\% from the corresponding
values obtained from the force scale
$r_0$~\cite{Guagnelli:1998ud}, which can
be explained by scaling violations.

\begin{figure}[t]
\vspace*{0.23cm}
\includegraphics[width=6.45cm]{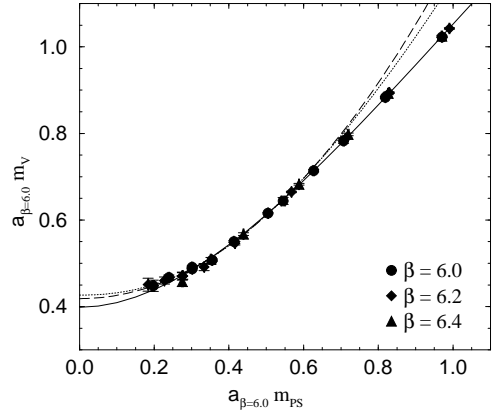}
\vspace*{-1.1cm}
\caption{Fits as in Fig.~\ref{fig:chiral_rho-lin} but with $C_{1/2}=0$.
         The dotted line comes from a fit using eq.~(\ref{eq:ansatz_pheno}).}
\vspace*{-0.6cm}
\label{fig:chiral_rho-nonlin}
\end{figure}

The fits shown in Fig.~\ref{fig:chiral_rho-lin} can be compared with
fits shown in Fig.~\ref{fig:chiral_rho-nonlin}, where $C_{1/2}$ has
been set to zero.
Also shown are the results for a phenomenological ansatz
\vspace*{-0.3cm}
\begin{equation}
(\tilde{a}_\beta m_{V,N})^2 = (\tilde{a}_\beta M_{V,N})^2 +
                      \sum_{i=2}^{3} A_i (\tilde{a}_\beta m_{PS})^i
\label{eq:ansatz_pheno}
\end{equation}
\vspace*{-0.1cm}
which has less free parameters.

\section{CONTINUUM LIMIT}

For improved fermions we expect scaling violations to be of
$O(a^2)$. We therefore extrapolate linearly in $(a/r_0)^2$.
To avoid artifacts of the quenched approximation, we use the
phenomenological ansatz
$(a m_{V,N})^2 = (a M_{V,N})^2 +
A_{2,\beta} (a m_{PS})^2 + A_{3,\beta} (a m_{PS})^3$
for the extrapolation to the chiral
limit, while we use a Pad\'e-like ansatz
similar to eq.~(\ref{eq:ansatz_pade})
with $C_{1/2}=0$ to
interpolate our data or to extrapolate
towards the charm quark mass region.
\begin{figure}[t]
\vspace*{0.2cm}
\includegraphics[width=6.6cm]{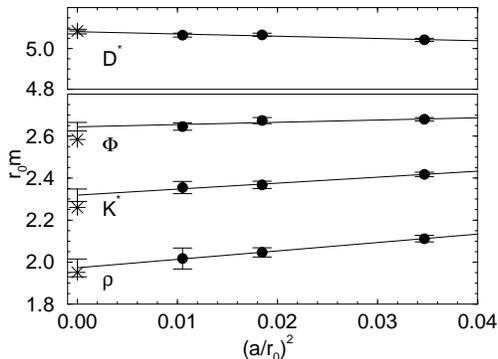}
\vspace*{-1.2cm}
\caption{Continuum extrapolation of the vector meson masses. The stars
         show the experimental value.}
\vspace*{-0.7cm}
\label{fig:spectrum_mv}
\end{figure}

\begin{figure}[t]
\includegraphics[width=6.60cm]{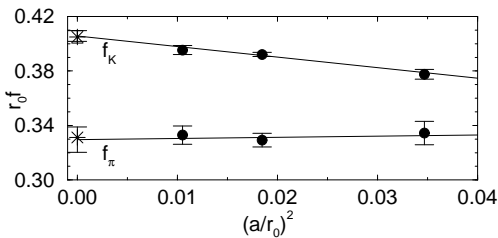}
\vspace*{-1.2cm}
\caption{The same as in Fig.~\ref{fig:spectrum_mv} for
the pseudoscalar decay constants
$f_\pi$ and $f_K$.}
\vspace*{-0.7cm}
\label{fig:spectrum_fps}
\end{figure}

\begin{figure}[t]
\vspace*{0.2cm}
\includegraphics[width=6.55cm]{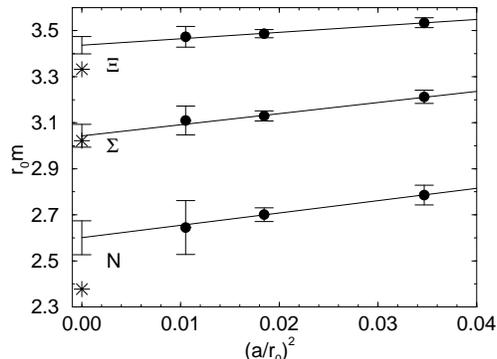}
\vspace*{-1.2cm}
\caption{The same as in Fig.~\ref{fig:spectrum_mv} for
spin-{\half} octet baryons.}
\vspace*{-0.9cm}
\label{fig:spectrum_baryons}
\end{figure}

In Fig.~\ref{fig:spectrum_mv} we show our results for the
vector meson masses. We used the experimental values for
the $\pi$, $K$, and $D$ mass to fix $\kappa_{u,d}$,
$\kappa_{s}$, and $\kappa_{ch}$, respectively. Since we use
results for degenerate quark masses, we have to
restrict ourselves to the light masses in the baryon sector.
The results are shown in Fig.~\ref{fig:spectrum_baryons}.

To one-loop {\qchpt}, the pseudoscalar decay constants with
degenerate quark masses should not suffer from quenched artifacts.
We again used the phenomenological ansatz to obtain the chiral limit.
The continuum extrapolation of $f_\pi$ and $f_K$ is shown in
Fig.~\ref{fig:spectrum_fps}.

\section{CONCLUSIONS}

We found evidence for non-analytic behavior of
$m_{PS}^2/m_q$ and $m_V$.
The results for $m_N$  do not seem to confirm the
predictions of {\qchpt}, but within errors we cannot conclude $C_{1/2}>0$.
Our result for the parameter $\delta$ agrees with the results of the
CP-PACS collaboration~\cite{Burkhalter:1998wu} and is slightly larger than
the number found by the FNAL group~\cite{Thacker:1998sb}. For $m_V$ we
found $C_{1/2}$ to be much larger than in~\cite{Burkhalter:1998wu}.

If we use the phenomenological ansatz for the extrapolation to the chiral
limit the ratio $m_N/m_\rho$ seems to become larger than the experimental
value. This is probably a quenching effect.
For all
masses and decay constants presented here discretization errors
are $O(a^2)$.

\vspace{-0.1cm}
\section*{ACKNOWLEDGEMENTS}

The numerical calculations were performed on the Quadrics {\it QH2} at DESY
(Zeuthen) as well as the Cray {\it T3E} at ZIB (Berlin) and the Cray
{\it T3E} at NIC (J\"ulich).


\begin{thebibliography}{9}

\bibitem{Sharpe:1993gr}
  S.R.~Sharpe,
  Nucl.\ Phys.\ B (Proc.~Suppl.) {\bf 30} (1993) 213.

\bibitem{Sint:1997jx_Capitani:1998mq}
  S.~Sint and P.~Weisz,
  Nucl.\ Phys.\ {\bf B502} (1997) 251;
  S.~Capitani {\it et al.},
  Nucl.\ Phys.\ {\bf B544} (1999) 669.

\bibitem{Booth:1997hk}
  M.~Booth, G.~Chiladze and A.F.~Falk,
  Phys.\ Rev.\ {\bf D55} (1997) 3092.

\bibitem{Labrenz:1996jy}
  J.N.~Labrenz and S.R.~Sharpe,
  Phys.\ Rev.\ {\bf D54} (1996) 4595.

\bibitem{Jaffe:1990jz}
  R.L.~Jaffe and A.~Manohar,
  Nucl.\ Phys.\ {\bf B337} (1990) 509.

\bibitem{Guagnelli:1998ud}
  M.~Guagnelli, R.~Sommer and H.~Wittig
  Nucl.\ Phys.\ {\bf B535} (1998) 389.

\bibitem{Burkhalter:1998wu}
  R.~Burkhalter {\it et al.},
  Nucl.\ Phys.\ B.\ (Proc.\ Suppl.) {\bf 73} (1999) 3.

\bibitem{Thacker:1998sb}
  H.~Thacker {\it et al.},
  Nucl.\ Phys.\ B (Proc.\ Suppl.) {\bf 73} (1999) 243.

\end{thebibliography}
\end{document}